\documentclass[a4paper,12pt]{article}
\usepackage{graphicx}
 
\begin{document}
\thispagestyle{empty}
\vspace*{-15mm}
%----------
\baselineskip 10pt
\begin{flushright}
\begin{tabular}{l}
{\bf }\\
{\bf }\\
{\bf }
\end{tabular}
\end{flushright}
\baselineskip 24pt 
\vglue 10mm 
%%%%%%%%%%%%%%%%%%%%%%%%%%%%%%%%%%%%%%%%%%%%%%%%%%%%%%%%%%
%%%%%%%%%       TITLE      %%%%%%%%%%%%%%%%%%%%%%%%%%%%%%%
%%%%%%%%%%%%%%%%%%%%%%%%%%%%%%%%%%%%%%%%%%%%%%%%%%%%%%%%%%
\begin{center}
{\Large\bf
Chaotic sources and Percolation of strings }
\vspace{8mm}

\baselineskip 18pt 
\def\thefootnote{\fnsymbol{footnote}}
\setcounter{footnote}{0}

{\bf M. A. Braun\footnote{E-mail: mijail@fpaxp1.usc.es.}}\\
\vspace{4mm}
{\it High-energy departament}\\
{\it St. Petersburg University}\\
{\it 198904 St. Petersburg, Russia}\\

\vspace{5mm}

{\bf F. del Moral\footnote{E-mail: felix@fpaxp1.usc.es} 
and C. Pajares\footnote{E-mail: pajares@fpaxp1.usc.es.}}\\
\vspace{4mm}
{\it Departamento de F\'{\i}sica de Part\'{\i}culas}\\
{\it Universidade Santiago de Compostela}\\
{\it 15706 Santiago de Compostela, Spain}\\

\vspace{10mm}
\end{center}
%%%%%%%%%%%%%%%%%%%%%%%%%%%%%%%%%%%%%%%%
%%%%%                              %%%%%
%%%%%          Abstract            %%%%%
%%%%%                              %%%%%
%%%%%%%%%%%%%%%%%%%%%%%%%%%%%%%%%%%%%%%%
\begin{center}
{\bf Abstract}\\[7mm]
\begin{minipage}{12cm}
\baselineskip=0.6cm
\noindent
%%%%%----------------------------------
It is shown that different ways of interacting strings formed in high energy nucleus-nucleus collisions cause
a different strength of the chaoticity parameter $\lambda$ of Bose-Einstein correlations. In particular, in the case
of percolation of strings, $\lambda$ shows a peculiar dependence on the string density, very similar to the 
dependence of the fractional average cluster size. In both, the derivative on the string density is maximum at
the critical point. The reasonable agreement with the existing experimental data indicates that percolation of 
strings can actually occurs.
%%%%%----------------------------------  
\end{minipage}
\end{center}

\newpage
\baselineskip=0.6cm 
%%%%%%%%%%%%%%%%%%%%%%%%%%%%%%%%%%%%%%%%%%%%%%%%%%%%%%%%%%%%%%
%%%%%%%         BEGINNING OF TEXT      %%%%%%%%%%%%%%%%%%%%%%%
%%%%%%%%%%%%%%%%%%%%%%%%%%%%%%%%%%%%%%%%%%%%%%%%%%%%%%%%%%%%%% 
In the search of the possible Quark-Gluon-Plasma (QGP) state of matter there are several interesting 
studies \cite{wiede,pratt}
on the space-time extension of the production region from which finally observed particles emerges, looking
at the Bose-Einstein correlation (BEC) for two identical particles
\begin{eqnarray} && C_2(Q^\mu)=1+ \lambda f(Q^\mu) \\
&& Q^\mu=p_1^\mu-p_2^\mu \qquad f(0)=1 \end{eqnarray}

It is well known that the shape of the function $f(Q^\mu)$ can give information on the size of the 
source \cite{pratt,weiner}. In
this paper we look at the correlation strength parameter $\lambda$, also called chaoticity, showing that also
bears valuable information related to the dynamics of the multiparticle production and in particular to the 
possible formation of QGP. In particular, as $\lambda$ counts the number of effective independent 
sources \cite{giovan,alexan}, 
it will behave in a very different way depending on how interact these sources and therefore of the resulting
number of effective independent sources.

We assume, as most of the models of soft hadronic interactions, that strings are formed among the partons
of projectile and the target of the collision. The fragmentation of the strings is due to the formation of 
quark-antiquark
or diquark-antidiquark pairs in different points. Each string can be considered as a totally chaotic source 
$\lambda=1$ \cite{anders} (The possibility of some degree of coherence also has been considered in some version of
fragmentation \cite{bowler}). Each string has a transverse size $\pi r^2$, determined by the colour field streched 
between the colour charges of the partons placed at the extremes of the strings. As the energy and/or 
the size of projectile and target increases the number and density of strings increases, originating
the interaction among strings which are not independent any longer. There are several possibilities of interaction.
We will considerer three main cases:

a) String fusion \cite{amelin} or colour rope \cite{sorge} formation. Strings fuse as soon 
as their transverse positions come within
a certain interaction area of the order of the string transverse dimension $\pi r^2$. The fusion of 
strings may take place only when their rapidity intervals overlap. The emerging string has the 
energy-momentum sum of the energy-momentum of the original strings and the same transverse size $\pi r^2$. The
colour properties of the formed string is determined from the standard SU(3)-colour composition laws. The 
transverse size of the new string is the same as the one of the original strings.

b) Clusters of strings where the total transverse size is the geometrical one. Clusters with the same number 
of strings can have different sizes depending on the way of overlapping of the strings. In this case we are
going to assume that the colour field is SU(3) summed only in the overlapping 
regions \cite{braun}. Therefore one cluster
should be considerer as several independent sources in most of the cases.

c) The clusters of strings have also the geometrical size but the colour-field is homogeneous all over
the cluster area as water drops \cite{armesto}. Each cluster can be considerer as a single source of incoherent
(chaotic) production of particles.

In case a) of string fusion or colour rope there is not possibility of phase 
transition \cite{braun2}. In cases b) and c) at
high energy and/or large projectile and target when the string density reach a critical value $\eta_c$ there
is at least one path formed of overlapping strings through the transverse area of the collision. A second
order phase transition takes place, the percolation of 
strings \cite{armesto2}. The critical value $\eta_c$ of $\eta$,
\begin{equation} \eta= \pi r^2 {N \over \pi R^2} \label{eta}\end{equation}

takes the value 1.17-1.5 depending on the profile functions of the colliding nuclei used. $N$ 
is the number of strings of transverse size $\pi r^2$ formed in a collision whose transverse area is $\pi R^2$.

Taking into account that each string is an incoherent source ($\lambda=1$), it can be shown \cite{biyajima}
\begin{equation}\lambda={n_S \over n_T}\label{lambda}\end{equation}

where $n_S$ are the number of pair particles produced in the same string in the rapidity and transverse 
momentum range fixed, and $n_T$ is the total number of particles produced in the same kinematic range. To
obtain equality (\ref{lambda}) it is neglected the effects of B-E correlations on the single inclusive cross section.
This fact, must be taken into account to correct the value of $\lambda$ obtained from formula (\ref{lambda}).
We correct this value in the same way a some experiments do \cite{na44}, obtaining a new value $\lambda$
\begin{equation}\lambda=K_{spc}(\lambda_u+1)-1\end{equation}

where $\lambda_u$ is the uncorrected one and $K_{spc}$ is the single particle correction which is evaluated
for each collision and kinematics range in the same way described in reference \cite{na44}. To obtain 
(\ref{lambda}) it is assumed that these are not BEC among pairs coming from different strings. This 
assumption is in agreement with L3, Delphi an ALEPH studies on BEC in $W^+W^-$ experimental events 
produced in $e^+e^-$ at LEP \cite{lep}.

Qualitatively, it is clear that without any kind of interactions of strings $\lambda=1$ and $\mu= n \mu_1$
where $N$, $\mu$ and $\mu_1$ are the total number of strings, the average total multiplicity and the mean
multiplicity of one string. In the case a) of string fusion, the dominant term is
\begin{equation}\lambda={1 \over <M>} \qquad <M>=< \sum_n \nu_n>\label{casea}\end{equation}

where $<M>$ is the mean value of resulting strings, (clusters), and $\nu_n$ is the number of clusters formed
from $n$ original strings. As the probability of obtaining $\nu_n$ clusters is \cite{braun2}
\begin{equation}P(\nu_n)=  {c\, p^{N-M} \over \prod_{n=1} (\nu_n! (n!)^{\nu_n})}
\prod_{k=1}^{M-1} (1-kp)\end{equation}

where $c$ is a normalization constant, and $p$ the fusion probability $p=  r^2 / R^2$
the mean values are
\begin{equation}<\nu_n>= C_N^n p^{n-1} (1-p)^{N-n} \quad,\quad <M>={1 \over p} (1-(1-p)^N)\end{equation}

In the {\emph {thermodynamical limit}} $p\rightarrow 0$, $N\rightarrow\infty$, the relevant parameter is
$\eta=N p$ and
\begin{equation}{<M>\over N}= {1 \over \eta}(1-e^{-\eta})= F(\eta)^2\end{equation}

therefore
\begin{equation}\lambda \rightarrow{1 \over N F(\eta)^2}\end{equation}

As $\mu= N F(\eta)\mu_1$, the quantity
\begin{equation}{\lambda\mu\over\mu_1}\rightarrow{1 \over   F(\eta) }\end{equation}

only depends on $\eta$.

In the case b), denoting by $n_i$ the number of regions where there are just overlapping of $i$ strings, $S_{ij}$
the area of the j-th region where there are just overlapping of $i$ strings, $S_i=\sum_{j=1}^{n_i} S_{ij}$
the total area where there are overlapping of just $i$ strings, $\sigma= \pi r^2$, and $\alpha=2\arccos (R/2)$,
we have in the same approximation as in (\ref{casea})
\begin{equation}\lambda= { <\sum_{i=1}^N \sum_{j=1}^{n_i} i {S_{ij}^2 \over \sigma^2}> \over 
<(\sum_{i=1}^N \sqrt{i} {S_i \over \sigma})^2>}\end{equation}

using the results of reference \cite{braun}, in the thermodynamical limit we have
\begin{eqnarray} \lambda &=& { I \over N F(\eta)^2 }\\
I &=&  \int_0^2 dR \, R {2 \over \pi} (\alpha - \sin \alpha) \exp \big \{ -2 \eta\,
[ 1-{1 \over \pi}(\alpha - \sin \alpha)]\big \}\end{eqnarray}

And so
\begin{equation}{\lambda\mu\over\mu_1} = {1 \over F(\eta) } I\end{equation}

depends only on $\eta$.

Finally in case c)
\begin{equation}\lambda= { <\sum_{i=1}^N  \nu_i (\sqrt{i})^2> \over 
<(\sum_{i=1}^N \nu_i \sqrt{i})^2>}= {N \over<(\sum_{i=1}^N \nu_i \sqrt{i})^2> }  \end{equation}

which scales on $\eta$, and $\lambda\rightarrow 1$ as $\eta\rightarrow\infty$.

\begin{figure}[!h]
 \begin{center}
  \leavevmode
    \includegraphics[scale=1.2]{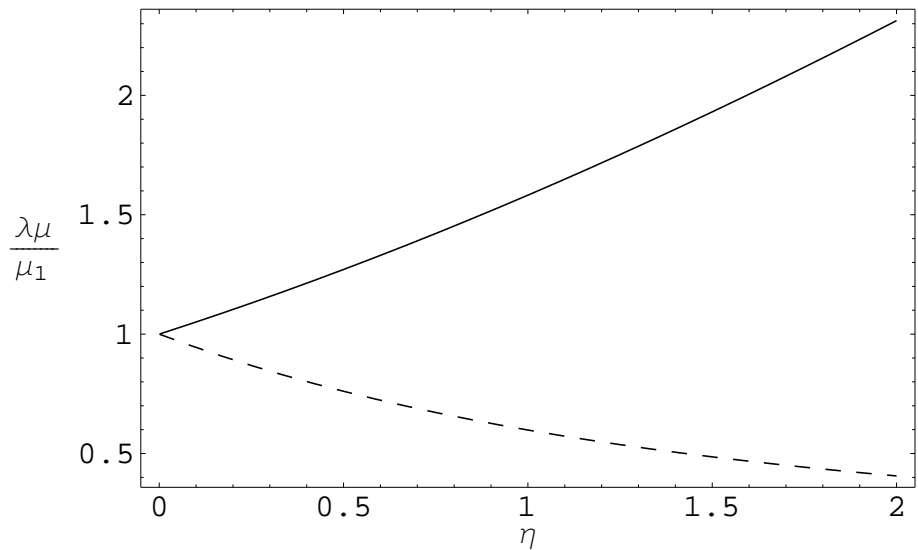}   
    \caption{ \small Dependence of the product of the chaoticity $\lambda$ and the ratio between
    multiplicities as a function of $\eta$ for string fusion (solid line) and case b) (dashed line).}
    \label{figure1}
 \end{center}
\end{figure}

\begin{figure}[!h]
 \begin{center}
  \leavevmode
    \includegraphics[scale=1.2]{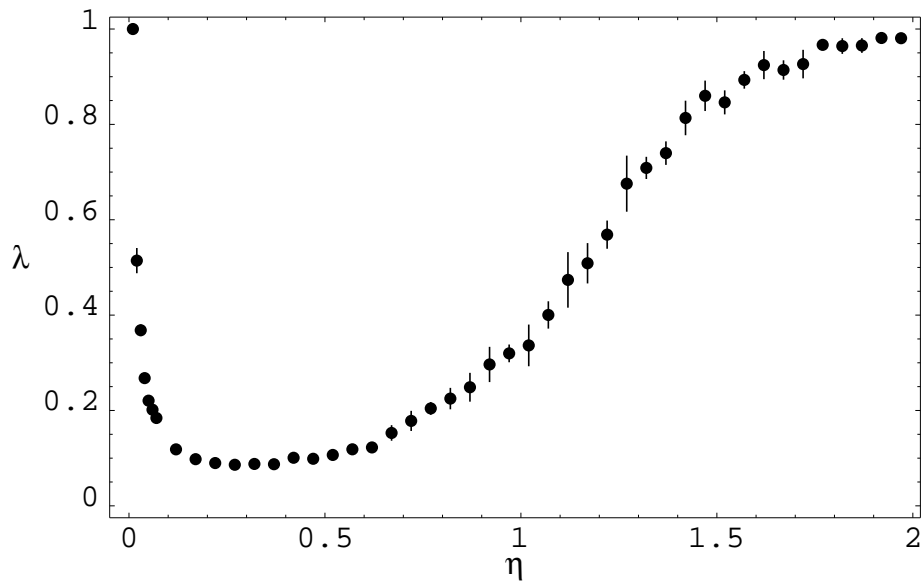}   
    \caption{ \small Dependence of $\lambda$ on $\eta$ for the case c) of percolation of strings.}
    \label{figure2}
 \end{center}
\end{figure}

In fig. \ref{figure1}, it is plotted ${\lambda\mu \over \mu_1}$ as a function of $\eta$ for cases a) and b). In
fig. \ref{figure2} it is shown the behavior of $\lambda$ on $\eta$ for the case c). It is shown that in this case
$\lambda$ drops as $\eta$ increases, for $\eta$ less than 0.25. For further increases of $\eta$, $\lambda$
increases even for $\eta$ less than the critical percolation threshold $\eta_c$. This behaviour is similar
to the average fractional cluster \cite{nardi} size as a function of $\eta$. Also ${d \lambda \over d \eta}$ vanishes
at $\eta\approx\eta_c$. This behavior, contrary to the cases a) and b) is in qualitative agreement with the
data as we will show below. To be more quantitative, we need to take into account energy-momentum conservations, 
and the energy-momentum and multiplicity of each cluster. We simulate the collisions using the framework of
the string fusion model code \cite{amelin}. In this code, it is know the transverse coordinates of the partons which form
each string, and their energy-momentum distribution. As the transverse size of each string $\pi r^2$, we can 
located in the impact parameter plane, each string and therefore to know the clusters formed and their 
energy-momentum. We know that in the case c) a cluster formed from $n$ strings on average will produce 
$\sqrt {n}$ times more particles than the produced in one string. This is the main information which we use in 
the simulation. The rapidity distribution of a decaying cluster formed with $n$ strings is taken a gaussian with
the proper normalization to take into account the mentioned factor $\sqrt {n}$. Instead gaussians we have
used other reasonable parametrizations without any strong change in our results.

\begin{figure}[!h]
 \begin{center}
  \leavevmode
    \includegraphics[scale=1.2]{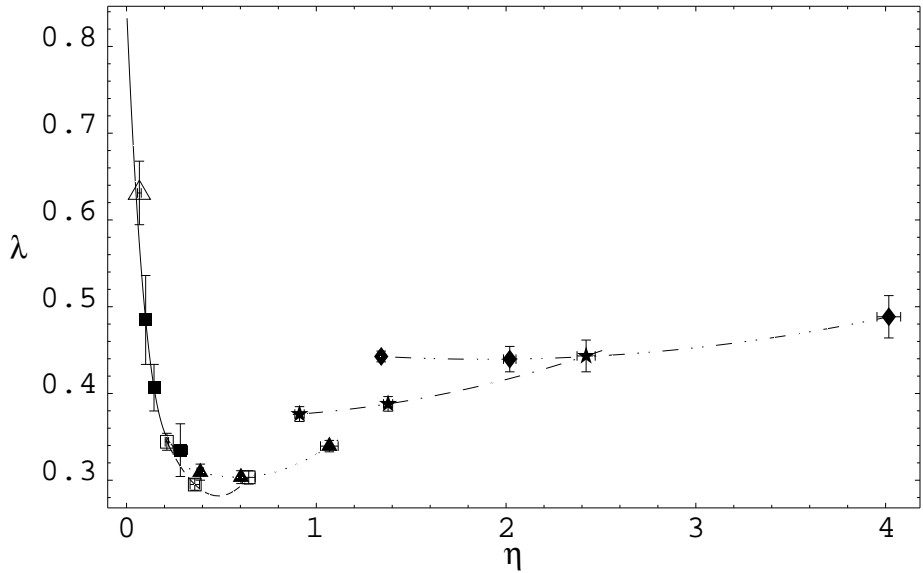}   
    \caption{ \small Dependence of $\lambda$ on $\eta$ for the case c) of percolation of strings in the
    MonteCarlo simulation. Points are $K_{spc}$ corrected and correlations are calculated between identical
    pions. Nonfilled triangle corresponds to C-C 
    minimum bias collisions at SPS energy; filled boxes to S-S minimum bias collisions at SPS, RHIC and
    LHC energies; nonfilled boxes to O-O central collisions ($b \leq 3.2$ fm.) at SPS, RHIC and
    LHC energies; filled triangles to S-S to central collisions ($b \leq 3.2$ fm.) at SPS, RHIC and
    LHC energies; stars to Ag-Ag central collisions ($b \leq 3.2$ fm.) at SPS, RHIC and LHC energies 
    and diamonds to Pb-Pb central collisions ($b \leq 3.2$ fm.) at SPS, RHIC and LHC energies.}
    \label{figure3}
 \end{center}
\end{figure}

In figure \ref{figure3} we present our results of $\lambda$ versus $\eta$ in the case c) evaluated for 
$y_{1cm}=y_{2cm}=0.5$. The error bars represents the uncertainties due to different parametrizations of
the string fragmentation.

It is seen that the shape is similar to the curve of figure \ref{figure2}. There are small violations of 
$\eta$ scaling due to minor differences which can appear in the detailed kinematics of the strings formed
in different nucleus-nucleus collisions, producing differences in the string density in the rapidity
interval considered, and also due to finite size effects. In anycase these differences are less than 10\%.
Our evaluations were done at central rapidity. Notice that the $\lambda$ value for a fixed collision depends
on the rapidity range studied. In the extreme of the rapidity range less strings are formed and a 
higher $\lambda$ is obtained.

The different experimental data are obtained in very different kinematics situations and also at very
different centrality of the collisions what does uneasy the comparison and even the comparison among
them. In anycase, data with ligth projectile as hadrons or Oxygen 
shows \cite{breaks,wa80} that $\lambda$ falls as
the multiplicity or the size of the target increases. For instance the values of $\lambda$ for
O-C, O-Cu, O-Ag and O-Au are \cite{wa802} 0.92, 0.29, 0.22 and 0.16 respectively (These numbers corresponds to a
centrality given by the average number of participants equal to 19.2, 39.5, 47.2 and 52.9 respectively.
The rapidity range considered was $-1\leq y_{lab}\leq1$). The same experiment show some weaker decrease for
different projectiles: $\lambda$=0.45, 0.32 and 0.33 for p, O and S respectively colliding against
Au target. NA44 quoted a value of 0.56 for S-Pb central collisions (3\% centrality) and 0.59 for
Pb-Pb central collisions (15\% centrality), (the range of pseudorapidity 
is $1.8< \eta<3.3$) \cite{na442}. In other
kinematic situation NA44 
quoted \cite{na44} $\lambda=0.46$ for S-Pb ($3.1< y<4.3$). NA49 obtains \cite{na49} $\lambda=0.42$ for
central Pb-Pb collisions ($2.9< y<5.5$, $p_T<0.6$).

The experimental situation on $\lambda$ is not clear but some trends can be distinguised. Fist, a falling
of $\lambda$ with the multiplicity for low density of collisions. Second, as the density becomes higer
$\lambda$ stop of decreasing and increases. This trend is just the behaviour obtained for $\lambda$ in
the case of percolation strings as water drops. The forthcoming experiments at RHIC and LHC can confirm
this behaviour by means of systematic studies of a range of projectiles and targets and for 
different centralities.

%%%%%%%%%%%%%%%%%%%%%%%%%%%%%%%%%%%%%%%%%%%%%%%%%%%%%%%%%%
%%%%%%%%%  ACKNOWLEDGEMENTS  %%%%%%%%%%%%%%%%%%%%%%%%%%%%%
%%%%%%%%%%%%%%%%%%%%%%%%%%%%%%%%%%%%%%%%%%%%%%%%%%%%%%%%%%
\vspace{0.5cm}
{\bf \Large Acknowledgments}
\vspace{0.3cm}
 
This work has been done under contract AEN99-0589-C02 of CICYT of Spain. F.d.M. thanks Xunta de
Galicia for a fellowship. We thank N. Armesto and D. Sousa for computation help and comments.
We thank J. Seixas for ask a question which originates this work.

%%%%%%%%%%%%%%%%%%%%%%%%%%%%%%%%%%%%%%%%%%%%%%%%%%%%%%%%%%
%%%%%%%%%  BIBLIOGRAPHY    %%%%%%%%%%%%%%%%%%%%%%%%%%%%%%%
%%%%%%%%%%%%%%%%%%%%%%%%%%%%%%%%%%%%%%%%%%%%%%%%%%%%%%%%%%

%%%%%%%%%%%%%%%%%%%%%%%%%%%%%%%%%%%%%%%%%%%%%%%%%%%%%%%%%%
%%%%%%%%%    LIST OF FIGURES  %%%%%%%%%%%%%%%%%%%%%%%%%%%%
%%%%%%%%%%%%%%%%%%%%%%%%%%%%%%%%%%%%%%%%%%%%%%%%%%%%%%%%%%
\newpage
\vspace{0.5cm}
{\bf \Large Figure captions}
\vspace{0.3cm}

{\bf Fig. 1.} Dependence of the product of the chaoticity $\lambda$ and the ratio between
    multiplicities as a function of $\eta$ for string fusion (solid line) and case b) (dashed line).

{\bf Fig. 2.} Dependence of $\lambda$ on $\eta$ for the case c) of percolation of strings.

{\bf Fig. 3.} Dependence of $\lambda$ on $\eta$ for the case c) of percolation of strings in the
    MonteCarlo simulation. Points are $K_{spc}$ corrected and correlations are calculated between identical
    pions. Nonfilled triangle corresponds to C-C 
    minimum bias collisions at SPS energy; filled boxes to S-S minimum bias collisions at SPS, RHIC and
    LHC energies; nonfilled boxes to O-O central collisions ($b \leq 3.2$ fm.) at SPS, RHIC and
    LHC energies; filled triangles to S-S to central collisions ($b \leq 3.2$ fm.) at SPS, RHIC and
    LHC energies; stars to Ag-Ag central collisions ($b \leq 3.2$ fm.) at SPS, RHIC and LHC energies 
    and diamonds to Pb-Pb central collisions ($b \leq 3.2$ fm.) at SPS, RHIC and LHC energies.

%%%%%%%%%%%%%%%%%%%%%%%%%%%%%%%%%%%%%%%%%%%%%%%%%%%%%%%%%%
%%%%%%%%%      FIGURES        %%%%%%%%%%%%%%%%%%%%%%%%%%%%
%%%%%%%%%%%%%%%%%%%%%%%%%%%%%%%%%%%%%%%%%%%%%%%%%%%%%%%%%%
\newpage

\end{document}